\documentclass[preprint,showpacs,preprintnumbers,amsmath,amssymb]{revtex4}

\usepackage{graphicx}% Include figure files
\usepackage{dcolumn}% Align table columns on decimal point
\usepackage{bm}% bold math

\begin{document}

\newcommand{\goo}{\,\raisebox{-.5ex}{$\stackrel{>}{\scriptstyle\sim}$}\,}
\newcommand{\loo}{\,\raisebox{-.5ex}{$\stackrel{<}{\scriptstyle\sim}$}\,}

%\begin{frontmatter}

\title{Production of exotic hypernuclei and hyper-matter}

\author{A.S.~Botvina}
%\author[INR]{A.S.~Botvina}
%\author[HIM,INR]{A.S.~Botvina\corauthref{cor1}}
%\author[IKP]{J.~Pochodzalla}

%   \address[HIM]{Helmholtz Institute Mainz, J. Gutenberg University, 
%55099 Mainz, Germany}

\affiliation{Institute for Nuclear 
Research, Russian Academy of Sciences, 117312 Moscow, Russia} 

%   \address[INR]{Institute for Nuclear Research, RAS, 
%117312 Moscow, Russia}

%   \address[IKP]{Johannes Gutenberg-Universit{\"a}t Mainz, Institut 
%f{\"u}r Kernphysik, D-55099 Germany}

%   \address[GSI]{GSI - Helmholtzzentrum f\"ur Schwerionenforschung GmbH, 
%62491 Darmstadt, Germany}

%\corauth[cor1]{Corresponding author: a.botvina@gsi.de}

%\date{\today}

\begin{abstract}
Pioneering experiments on production of hypernuclei can be performed 
with nuclotron beams on fixed targets, and at the future NICA facility. 
The peripheral collisions of relativistic ions are very promising 
for searching mutli-strange and exotic 
hypernuclei which are not easy accessible with other experimental methods. 
In these experiments one can also get information on the Equation of State 
of hyper-matter around nuclear saturation density at low and moderate 
temperatures. 
\end{abstract}

\pacs{25.75.-q , 21.80.+a , 25.70.Mn }

\maketitle

%\end{frontmatter}

% Proposal of new experiments with the nuclotron beams at JINR (Dubna),
% submitted to the NICA White Book. 

%\vspace{0.2cm}

%{\large PACS: 25.75.-q , 21.80.+a , 25.70.Mn }

%\vspace{0.5cm}

In nuclear reactions at high energies strange particles (baryons and 
mesons) are produced abundantly, and they are strongly involved in 
the reaction process. 
The specifics of hypernuclear physics is that there is no direct
experimental way to study hyperon--nucleon ($YN$) and hyperon--hyperon
($YY$) interactions ($Y=\Lambda,\Sigma,\Xi,\Omega$). 
When hyperons are captured by nuclei, hypernuclei are produced, 
which can live long enough in comparison with nuclear reaction times. 
Therefore, a nucleus may serve as a laboratory offering a unique opportunity 
to study basic properties of hyperons and their interactions.
Double- and multi-strange nuclei are especially interesting, because
they are more suitable for extracting information about the hyperon--hyperon
interaction and strange matter properties.  

The investigation of hypernuclei allows to answer many fundamental 
questions: Studying the structure of hypernuclei helps to understand the 
structure of conventional nuclei too \cite{japan} and it leads to an 
extension of the nuclear chart into the strangeness sector 
\cite{cgreiner}. Hypernuclei 
provide a bridge between traditional nuclear physics (dealing with 
protons and neutrons) and hadron physics. Strangeness is an important 
degree of freedom for the construction of QCD motivated models of strong 
interactions \cite{schramm}. Hyperons are also very important in many 
astrophysical sites, e.g., they are abundantly produced in nuclear 
matter at high densities, which are realized in the core of neutron 
stars \cite{schaffner}. 
The only way to describe realistically these physical conditions is to 
study the hyperon interactions in laboratory, and select theoretical models 
which pass the careful comparison with experimental data. 

It has been realized that the absorption of hyperons in spectator regions 
of peripheral relativistic ion collisions is a promising way for producing 
hypernuclei \cite{wakai1,cassing,giessen,botvina2011}. 
The basic mechanisms of such reactions were well established in analysis 
of collisions of conventional nuclei: Nucleons in the overlapping zone 
between the projectile and the target (participants) interact intensively 
with each other and produce many new particles including strange ones. These 
particles can re-scatter and propagate further towards the non-overlapping 
parts of nuclei (spectator residues) and can be captured there, if their 
relative kinetic energy is smaller than their potential in nuclear matter. 
This mechanism will lead to formation of excited normal spectators and
hyper-spectators, which later on disintegrate into ordinary and 
hyper-fragments \cite{bot-poch}. 
Very peripheral collisions lead to spallation of the normal nuclei and to 
fission of heavy ones. In mid-peripheral collisions, when temperatures 
of excited spectators reach several MeV, 
multifragmentation reaction takes place, and this allows for investigation 
of the Equation of State (EOS) in the region of the nuclear liquid-gas 
phase transition \cite{smm,ogul}. Since the $\Lambda$ potential in nuclear 
matter is of the same order as the nucleon one we expect that similar 
processes will occur with hyper-spectators too. 
Such reactions may give access to heavy and exotic 
hypernuclei, as well as to very strange nuclei beyond S=--2 
\cite{botvina2011}. It was also predicted that the relative yields of 
hypernuclei produced in these reactions can reveal important information 
about their properties and provide a unique way for experimental 
studying hyper-matter at relatively low temperatures ($T \loo 10$ MeV) 
\cite{bot-poch}. These are important advantages of the proposed 
measurements over experimental methods used presently in hyper-physics, 
which are mostly limited by investigations of hypernuclei with S=--1 
in ground and weakly-excited states. 

Actually, early experiments with 
light-ion beams at the LBL~\cite{Nie76} and JINR~\cite{Avr88} 
have demonstrated that hypernuclei can be formed in such reactions. 
Recently the HypHI collaboration at GSI Darmstadt has reported 
first results on the production of light hypernuclei in the disintegration 
of 2 GeV per nucleon $^6$Li projectiles impinging on a $^{12}$C target 
\cite{saito-new}. 
This experiment has confirmed the feasibility to produce hypernuclei 
abundantly in peripheral ion collisions. 
The observed production of $^{3}_{\Lambda}$H is by about a factor of 
three larger than the production of $^{4}_{\Lambda}$H. In a still 
preliminary analysis also indications for a significant 
bump in the $\pi^{-}$-deuteron 
invariant mass distribution were found \cite{saito}. If confirmed in the 
ongoing analysis this observation could be interpreted as the formation of 
slightly bound $\Lambda$-neutron systems ($\Lambda n$ and $\Lambda n n$), 
which seem to dominate over other hypernuclei. 

It was theoretically demonstrated that this reaction can be explained 
within a hybrid approach including dynamical stage of production and 
capture of hyperons by spectators and statistical stage describing 
decay of such excited spectators into hypernuclei 
\cite{bot2012,botvina2012}. 
The first stage of this collision 
process was described within the transport Dubna cascade model (DCM)
\cite{toneev83,toneev90} taking into account absorption of $\Lambda$-hyperons 
\cite{botvina2011}. The generalized for hyper-nuclei Fermi-break-up model 
\cite{lorente} was used for the second stage. 
\begin{figure}[t] 
\includegraphics[width=0.8\linewidth]{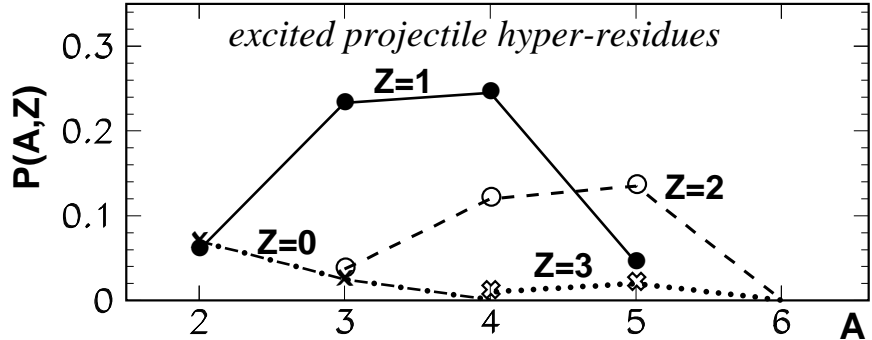} 
\caption{\small{
Ensemble of the projectile spectators with one absorbed $\Lambda$ 
hyperon as predicted by DCM calculations for $^6$Li (2 A GeV) + $^{12}$C 
reaction: P(A,Z) is relative probability to produce specific excited 
hyper-residues with mass A and charge Z. The dot-dashed, solid, dashed, and 
dotted lines correspond to residues with Z=0, 1, 2, and 3, respectively. 
}}
\label{fig1}
\end{figure}
The DCM calculations of $^6$Li (2 A GeV) + $^{12}$C collisions predict 
formation of a broad ensemble of projectile residues with captured 
$\Lambda$ hyperons. After integration over all impact parameters 
this ensemble is shown in Fig.~\ref{fig1}. 
The following evolution depends on excitation energy of these residues: 
Their baryon content will not change practically if the excitations are low. 
In the case of high excitations the residues will loose 
nucleons and small final hypernuclei will be produced predominantly, 
including exotic $\Lambda n$ hyper-systems \cite{botvina2012}. 

We emphasize specially that the exotic $\Lambda$--neutron systems were 
never observed previously with traditional methods of hyper-physics, which 
use mainly a capture of hyperons produced by hadrons and leptons of 
high energy in nuclei without their excitation. The 
reason may be that a very low binding energy expected for the $\Lambda n$ 
systems (around few tens keV) can not be seen in such direct interactions 
releasing particles with large energy. 
On the other hand, the decay of 
moderately excited hyper-systems into small hypernuclei \cite{lorente} 
is rather sensitive to the tiny binding energy. Therefore, the new 
reaction mechanism realized in peripheral collisions of relativistic 
ions makes possible to produce and investigate exotic hypernuclear 
species. 

\begin{figure}[tbh]
\includegraphics[width=0.8\linewidth]{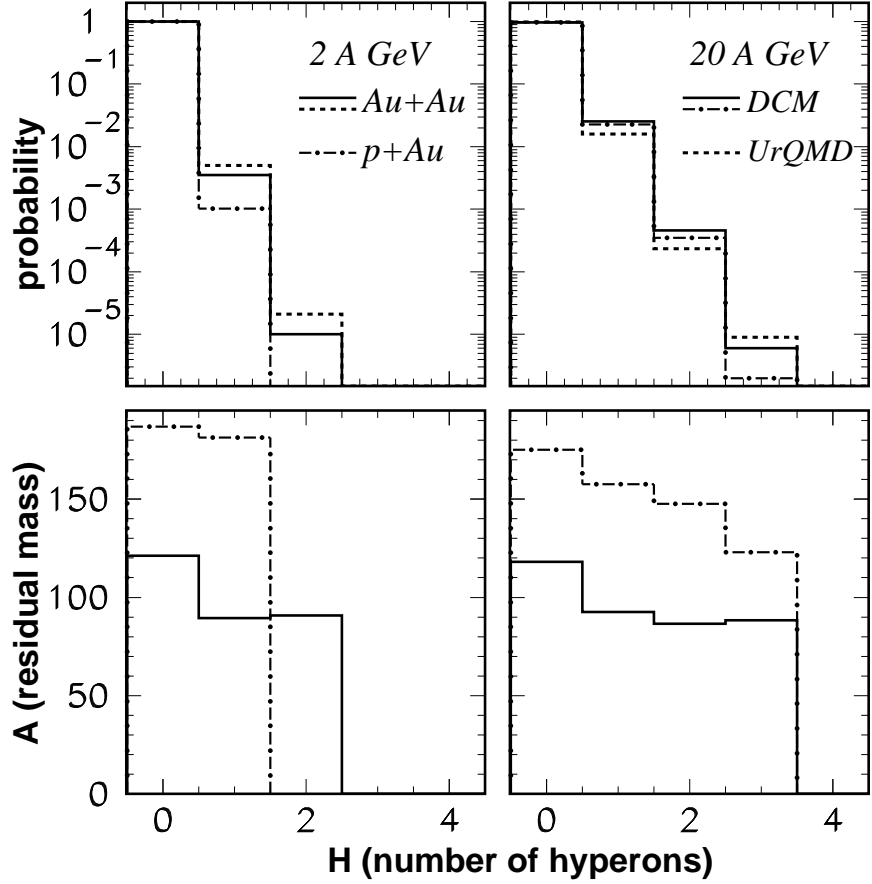}
\caption{\small{
Probability for formation of conventional and strange spectator residuals 
(top panels), and their mean mass numbers (bottom panels)
vs the number of captured $\Lambda$ hyperons (H), calculated with DCM 
and UrQMD model for p + Au and  Au + Au collisions with energy of 
2 GeV per nucleon (left panels), and 20 GeV per nucleon (right panels) 
\cite{botvina2011}. 
The reactions and energies are noted in the figure by different 
histograms. 
}}
\label{fig2}
\end{figure}

Another new important possibility is to study fragmentation and 
multifragmentation of hyper-matter produced in peripheral collisions 
of heavy ions. In this case one can address the EOS of hyper-matter 
at moderate temperatures as previously it was done for conventional 
matter \cite{ttt}: The method is to analyze the yields of fragments and 
hyper-fragments and their velocity correlations, which contain information 
about hyperon interaction in medium. For illustration, Fig.~\ref{fig2} shows 
probabilities for producing spectator residues with different 
numbers of captured $\Lambda$s, in collisions of proton on gold and 
gold on gold at energies 2 and 20 GeV per nucleons \cite{botvina2011}. 
The dynamical stage of these reactions was described by DCM 
\cite{toneev83,toneev90} and UrQMD (Ultra-relativistic 
Quantum Molecular Dynamics) \cite{Bleicher:1999xi,Bass:1998ca} models. 
One can see that both in proton-nucleus 
and nucleus-nucleus collisions the hyper-spectators can be formed 
abundantly. For example, the fraction of excited spectator 
residues with one $\Lambda$ is in the range of from 0.1\% (for the 
case of 2 GeV proton) to few percent of the total yield at 20 GeV. 
At the nuclotron's beam energy of 4.5 A GeV the estimated probability 
of one $\Lambda$ capture 
will be around 0.3\% for p + Au and more than 1\% for Au + Au collisions. 
The probabilities for capturing two $\Lambda$s will be by nearly two 
order lower, and the probability for three $\Lambda$ to be captured is 
around 10$^{-6}$. However, these reaction probabilities are quite 
sufficient for hyper-physical experiments which are usually adjusted for 
cross-sections about few nanobarns. The absorption of a 
higher number of hyperons is also feasible. This new mechanism opens a 
unique opportunity to produce and study multi-strange systems, which are not 
conceivable in other nuclear reactions. 
As discussed, later on the hot spectators disintegrate, and final 
fragments with products of weak decay of hypernuclei can be measured by 
detectors. Statistical models, see refs.~\cite{bot-poch,smm,lorente}, can be 
used to describe this stage of the process. 
Relativistic spectators have also obvious experimental advantages: 
Because of the Lorentz 
factor their lifetime is longer, the projectile hyper-fragments can 
travel a longer distance. This makes possible to use sophisticated vertex 
detectors and fragment separation technique for their identification. 

One should remember, that 
the considered reaction process is qualitatively different from the 
production mechanisms of light hypernuclei coming from decay of very 
excited (T$\sim 160$ MeV) fireballs with strangeness admixture, which are 
formed in central 
relativistic heavy-ion collisions, . In this case coalescence-like processes 
are most probable for cluster production. There is hard to expect 
production of big and weakly bound nuclei, because of a large energy 
deposited in the fireball \cite{andronic,stein2}. Nevertheless, after full 
construction of the NICA facility, this kind of measurements may also be done 
by colliding relativistic beams and detecting species coming from the 
midrapidity region. 

In conclusion, an advantage of the proposed novel experiments is that 
they can be already performed at the first stage of 
the NICA project with nuclotron beams by using the fixed targets. 
It is instructive that the very first experimental identification 
of a hypernucleus was performed in similar collisions: This 
event was observed in a multifragmentation reaction induced by a cosmic 
energetic proton or ion in photo-emulsion \cite{danysz}. Recently, 
encouraging results in studying spectator hypernuclei were obtained in GSI 
experiments \cite{saito-new,saito}. As we have shown theoretically 
\cite{botvina2011,bot-poch,bot2012,botvina2012,lorente}, the process of 
formation of moderately-excited hyper-nuclear systems ($T \approx 1-10$ MeV)
in peripheral collisions of relativistic ions with their subsequent 
disintegration is a natural way to produce new and exotic hypernuclei. 
We know also from studies of conventional nuclei \cite{smm,ttt} that 
investigation of such systems can 
provide effective methods to extract the EOS of hyper-matter at 
densities around the nuclear saturation density. 
Hypernuclear physics will benefit strongly from 
exploring the new production mechanism realized in peripheral ion collisions 
and the novel detection technique 
associated with fragmentation reactions of excited nuclei.

\end{document}